\documentclass[12pt,preprint]{aastex}
\usepackage{subfigure}

\newcommand\ltsima{$\; \buildrel <\over\sim \;$}
\newcommand\simlt{\lower.5ex\hbox{\ltsima}}
\newcommand\gtsima{$\; \buildrel >\over\sim \;$}
\newcommand\simgt{\lower.5ex\hbox{\gtsima}}

\begin{document}


\title{MOA-2008-BLG-379Lb:
A Massive Planet \\ 
from a High Magnification Event with a Faint Source}


\author{D. Suzuki\altaffilmark{1,A}, A. Udalski\altaffilmark{2,B}, T. Sumi\altaffilmark{1,A}, D.P. Bennett\altaffilmark{3,A}, I.A. Bond\altaffilmark{4,A} \\
and \\
F. Abe\altaffilmark{5}, C.S. Botzler\altaffilmark{6}, M. Freeman\altaffilmark{6}, M. Fukagawa\altaffilmark{1}, A. Fukui\altaffilmark{7}, K. Furusawa\altaffilmark{5}, Y. Itow\altaffilmark{5}, C.H. Ling\altaffilmark{4}, K. Masuda\altaffilmark{5}, Y. Matsubara\altaffilmark{5}, Y. Muraki\altaffilmark{8}, K. Ohnishi\altaffilmark{9}, N. Rattenbury\altaffilmark{6}, To. Saito\altaffilmark{10}, H. Shibai\altaffilmark{1}, D.J. Sullivan\altaffilmark{11}, K. Suzuki\altaffilmark{5}, W.L. Sweatman\altaffilmark{4}, S. Takino\altaffilmark{3}, P.J. Tristram\altaffilmark{12}, K. Wada\altaffilmark{1}, and P.C.M. Yock\altaffilmark{6}\\ (MOA Collaboration) \\
M.K. Szyma\'{n}ski\altaffilmark{2}, M. Kubiak\altaffilmark{2}, I. Soszy\'{n}ski\altaffilmark{2}, G. Pietrzy\'{n}ski\altaffilmark{2,13}, R. Poleski\altaffilmark{2, 14}, K. Ulaczyk\altaffilmark{2}, and \L. Wyrzykowski\altaffilmark{2,15}\\ (OGLE Collaboration)}

\altaffiltext{1}{Department of Earth and Space Science, Graduate School of Science, Osaka University, 1-1 Machikaneyama, Toyonaka, Osaka 560-0043, Japan}
\altaffiltext{2}{Warsaw University Observatory, Al. Ujazdowskie 4, 00-478 Warszawa, Poland}
\altaffiltext{3}{Department of Physics, University of Notre Dame, Notre Dame, IN 46556, USA}
\altaffiltext{4}{Institute of Information and Mathematical Sciences, Massey University, Private Bag 102-904, North Shore Mail Centre, Auckland, New Zealand}
\altaffiltext{5}{Solar-Terrestrial Environment Laboratory, Nagoya University, Furo-cho, Chikusa, Nagoya, Aichi 464-8601, Japan}
\altaffiltext{6}{Department of Physics, University of Auckland, Private Bag 92019, Auckland, New Zealand}
\altaffiltext{7}{Okayama Astrophysical Observatory, National Astronomical Observatory, 3037-5 Honjo, Kamogata, Asakuchi, Okayama 719-0232, Japan}
\altaffiltext{8}{Department of Physics, Konan University, Nishiokamoto 8-9-1, Kobe 658-8501, Japan}
\altaffiltext{9}{Nagano National College of Technology, Nagano 381-8550, Japan}
\altaffiltext{10}{Tokyo Metropolitan College of Industrial Technology, Tokyo 116-8523, Japan}
\altaffiltext{11}{School of Chemical and Physical Sciences, Victoria University, Wellington, New Zealand}
\altaffiltext{12}{Mt. John University Observatory, University of Canterbury, P.O. Box 56, Lake Tekapo 8770, New Zealand}
\altaffiltext{13}{Universidad de Concepci\'on, Departamento de Astronomia, Casilla 160-C, Concepci\'on, Chile}
\altaffiltext{14}{Department of Astronomy, The Ohio State University, 140 West 18th Avenue, Columbus, OH 43210, USA}
\altaffiltext{15}{Institute of Astronomy, University of Cambridge, Madingley Road, Cambridge CB3 0HA, UK}
\altaffiltext{A}{Microlensing Observations in Astrophysics (MOA) Collaboration}
\altaffiltext{B}{Optical Gravitational Lensing Experiment (OGLE) Collaboration}


\begin{abstract}
We report on the analysis of high microlensing event MOA-2008-BLG-379,
which has a strong microlensing anomaly at its peak, due to a massive
planet with a mass ratio of $q = 6.9 \times10^{-3}$.
Because the faint source star crosses the large resonant caustic
, the planetary signal dominates the light curve. This is unusual for planetary microlensing
events, and as a result, the planetary nature of this light curve was not immediately
noticed. The planetary nature of the event was found when the Microlensing Observations in Astrophysics (MOA) Collaboration 
conducted a systematic study of binary microlensing events previously
identified by the MOA alert system.
We have conducted a Bayesian analysis based on a standard Galactic model
to estimate the physical parameters of the lens system. This yields
a host star mass of $M_{\rm L} = 0.56_{-0.27}^{+0.24}\ M_\odot$ orbited by a planet 
of mass $m_{\rm P} = 4.1_{-1.9}^{+1.7}\ M_{\rm Jup}$ at an orbital separation of
$a = 3.3_{-1.2}^{+1.7}$ AU at a distance of  $D_{\rm L} = 3.3_{-1.2}^{+1.3}$~kpc.
The faint source magnitude of $I_{\rm S} = 21.30$ and relatively
high lens-source relative proper motion of $\mu_{\rm rel} = 7.6 \pm 1.6~{\rm mas}~{\rm yr}^{-1}$
implies that high angular resolution adaptive optics or $\it Hubble\ Space\ Telescope$
observations are likely to be able to detect the source star, which would
determine the masses and distance of the planet and its host star.
\end{abstract}


\keywords{gravitational lensing, planetary systems}



\section{Introduction}

Since the first discovery of an extrasolar planet orbiting an ordinary star
in 1995 \citep{mq95}, more than 1000 planets have been found to date.
Most of the planets discovered were through the radial velocity  \citep{but06, bon11} 
and transit \citep{bor11} methods, which are the most sensitive to the planets 
orbiting close to their host stars. In contrast, direct imaging is the most sensitive
to young planets in orbits wider than that of Saturn \citep{hr8799}.
The gravitational microlensing method has unique sensitivity to the planets 
orbiting just outside of the snow line with masses down to Earth mass \citep{bennett96}.
The snow line \citep{il05, lecar06, ken06, kk08} plays a critical role in the core accretion 
theory of planet formation. Ices, including water ice, can condense beyond the snow line, 
which increases the density of solid material in the proto-planetary disk by a factor 
of a few, compared to the disk just inside the snow line. The higher density of solids
means that the solid planetary cores can form more quickly beyond the snow line,
so microlensing is unique in its ability to probe the planetary mass function
at the separation where planet formation is thought to be the most efficient.

To date, the primary contribution of the microlensing method to the study of exoplanets
is the statistical results indicating that cold planets in wide orbits are quite common
\citep{sumi10, gould10, cssn12}. Relatively low-mass cold Neptunes or super-Earths
are found to be more common than gas giants, which is in rough agreement with the core
accretion theory expectation. Microlensing is also unique in its sensitivity to planets
orbiting stars that are too faint to detect \citep{bennett_rev, gaudi_araa}. In fact, planetary mass 
objects can be detected even in cases  when there is no indication of a host star \citep{sumi11}.

The microlensing events that are monitored for the signals of extrasolar planets
are mainly identified by the two main microlensing survey groups,
the Microlensing Observations in Astrophysics (MOA; Bond et al. 2001; 
Sumi et al. 2003) and the Optical Gravitational Lensing Experiment (OGLE; Udalski 2003).
The MOA survey uses the MOA-II 1.8m telescope with a 2.2 ${\rm deg}^{2}$ 
field-of-view (FOV) CCD camera MOA-cam3 \citep{sako08} at Mt. John University 
Observatory in New Zealand. With this large FOV camera, MOA is able to observe 
$>40 {\rm deg}^{2}$ of the central Galactic Bulge fields every hour, with a higher
cadence for the fields with the largest microlensing rate.
MOA identifies about 600 microlensing events in progress each year and issues
public alert messages as each new event is identified.
The OGLE survey has been conducted with the 1.3m Warsaw Telescope at the 
Las Campanas Observatory in Chile, which is currently equipped with a 
1.4$\,{\rm deg}^{2}$ CCD mosaic camera. However, in 2008, when the microlensing
event MOA-2008-BLG-379 took place, OGLE was operating the OGLE-III survey
using a camera with a 0.35$\,{\rm deg}^{2}$ FOV. OGLE-III typically identified $\sim 100$
more events than MOA-II, but observed at a lower cadence with most fields only
being observed once per night. Currently, the OGLE-IV survey operates at a relatively
high cadence in the central Galactic bulge, but covers a much wider area of sky
at a lower cadence, which yields nearly 2000 microlensing event discoveries per
year. Most recently, the Wise microlensing survey \citep{wise12} has conducted high cadence
observations using a 1.0$\,{\rm deg}^{2}$ CCD mosaic camera on the 1.0m Wise
Telescope in Israel. They fill the gap in longitude between MOA and OGLE, but
due to their Northerly location, they can only observe the Galactic bulge for
a limited amount of time each night.

Microlensing follow-up groups, such as $\mu$FUN, PLANET, RoboNet, and 
MINDSTEp, observe microlensing events that have been previously identified
by one or more of the survey groups. They aim to ensure dense, 24 hr light
curve coverage on the highest priority microlensing events, and they contribute to
planet discoveries in a number of ways. They can identify planetary signals in
some of the microlensing events with relatively bright source stars
\citep{ogle390,batista11}, densely monitor high magnification microlensing
events \citep{udalski05,gould10,miyake11,bachelet12,yee12,han13}
that are known to have high sensitivity to planets \citep{griest98,rhie00}, or follow-up
on events once potential planetary anomalies have been found by the survey teams
\citep{sumi10,mrk11,kains13}. In addition, some events have been identified based
on higher cadence observations by the survey groups prompted by the real-time
identification of potential planetary anomalies \citep{bond04}.
However, if the survey cadence is high enough, it is possible to identify planets
from survey observations alone without the benefit of follow-up observations
from either the survey or follow-up teams. The event we present here, 
MOA-2008-BLG-379, is the planetary microlensing event that was found based
on survey observations alone, similar to
MOA-2007-BLG-192 \citep{ben08}, MOA-bin-1 \citep{bennett12} and MOA-2011-BLG-322 \citep{yossi13}.
We should note that there are many more planetary events,
such as MOA-2011-BLG-293 \citep{yee12} and OGLE-2012-BLG-0406 \citep{poleski13},
that could have been discovered based upon survey observations alone. 
And in fact, the primary reason for the lack of follow-up observations for 
MOA-2007-BLG-192 and MOA-2008-BLG-379 is the lack of experience in the 
microlensing community with the full range of possible planetary signals.

In this paper, we report on the analysis of a planetary microlensing event MOA-2008-BLG-379.
Section~\ref{sec-obs} describes the observations of this event, and 
Section~\ref{sec-data_red_mod} describes the reduction of the data and the 
light curve modeling.
In Section~\ref{sec-color}, source magnitude and color measurement using 
color-magnitude diagram (CMD) is discussed, and the derived angular Einstein 
radius is discussed in Section~\ref{sec-thetaE}.
In Section~\ref{sec-bayes}, we use a Bayesian analysis based on a standard
Galactic model to estimate the masses 
and distances of the lens system based on the angular Einstein radius.
Discussion and our conclusions are presented in Section~\ref{sec-conclude}.

\section{Observations}
\label{sec-obs}

Prior to 2009, the MOA-II observing strategy called for the observations of the
two fields with the highest lensing rate every 10 minutes and observations of the
remaining 20 fields every hour using the custom wide-band MOA-Red filter, which is 
roughly the sum of the Cousins $R$- and $I$-bands.
MOA-2008-BLG-379 was discovered in field gb8, which was observed with an hourly cadence
in 2006--2008.

The microlensing event MOA-2008-BLG-379 was identified by the MOA 
alert system \citep{bond01} at 
$\rm(R.A., decl.)(J2000)$ = ($17^{\rm h}58^{\rm m}49^{\rm s}.44$, -$30^{\circ}11'48''.95$), 
and was announced by the MOA collaboration at UT 22:00, 2008 August 9, 
(or ${\rm HJD}^\prime = 4688.42$). As can be seen from the light curve in Figure~\ref{light curve},
this was after the first caustic entrance and exit and the central cusp approach feature.
Given the sparse coverage and lack of OGLE data at this time, it was not immediately
obvious that this event was anomalous.
Two weeks later, at UT 20:00 2008 August 23, the event 
was also identified and announced by the OGLE Early Warning System 
(EWS; \citealt{ogle-ews}) as OGLE-2008-BLG-570. The delay in the identification of this
event by OGLE was due to the fact that the source was faint and was not located close to
the location of a ``star" that was identified in the OGLE reference image. This is
fairly common, as most bulge main sequence stars are not individually resolved at
$\sim 1^{\prime\prime}$ resolution. The source star, MOA-2008-BLG-379S, happens
to be located at an unusually large distance from the nearest star in the OGLE catalog,
and as a result, it could only be found via the ``new object" channel of the OGLE EWS.
In 2008 this channel was not run as often as the regular ``resolved star" channel
of the OGLE EWS. As a result, both the MOA and OGLE collaboration 
groups observed this event with normal cadence.

\section{Data Reduction and Modeling}
\label{sec-data_red_mod}

The MOA and OGLE data for MOA-2008-BLG-379 were reduced using the
respective MOA \citep{bond01} and OGLE \citep{udal03} photometry pipelines.
The initial reductions used the normal photometry available on the respective
MOA and OGLE alert web sites, and these data were used to show that the correct
model involved a planet. However, it was possible to obtain improved photometry
for both the MOA and OGLE data sets. The improved MOA photometry used
``cameo" sub-images centered on the target, and the improved OGLE photometry
was redone with source position determined from difference images
and a careful selection of reference images.

Both the OGLE and MOA photometry were found to have systematic errors
at the beginning and/or end of each observing season when observations are
only possible at high airmass. This may be due to low-level variability for a very
bright star about 3 arc seconds from the source, MOA-2008-BLG-379S.
So we removed the MOA data prior to ${\rm HJD}^\prime = {\rm HJD} - 2450000 = 4543$,
leaving 951 MOA data points in the range $ 4543 <  {\rm HJD}^\prime < 4762 $. 
The OGLE data consist of 294 $I$-band observations and 7 $V$-band observations,
as shown in Table \ref{info}. 

The error bar estimates from the photometry codes are normally accurate to a factor 
of 2 or better, and they provide a good estimate of the relative photometry errors for
each data set. These are sufficient to find the best light curve model, but in order to
estimate the fit parameter uncertainties, we need more accurate error bars, which 
have $\chi^2/{\rm dof} \simeq 1$ for each data set. Therefore, once we have found
the best fit model with the unmodified error bars, 
we modify the error bars with the formula
\begin{equation}
\sigma^\prime_i=k\sqrt{\sigma^2_i+e^2_{{\rm min}}}
\end{equation}
where $\sigma_i$ is the input error bar estimate (in magnitudes) for the $i$th data point, 
$k$ is the normalization factor, and $e_{\rm min}$ is the minimum error. The
modified error bars, $\sigma^\prime_i$, are used for all subsequent modeling and
parameter uncertainty determination.

The factors $k$ and $e_{\rm min}$ are estimated for each data set with the following
procedure. First, we plot the cumulative distribution of $\chi^2$ as a function of the size of 
the input error bars, $\sigma_i$.  Then, we chose the value of $e_{\rm min}\geq 0$ 
such that the cumulative distribution is a straight line with slope of unity. Then,
the parameter $k$ is chosen to give $\chi^2/{\rm dof}\simeq1$ for each data set.
For this event, we find that $e_{\rm min}=0$ for all data sets. The values of
$k$ and $e_{\rm min}$ for all three data sets are listed in Table \ref{info}.

This event was identified in a systematic modeling of all anomalous events (i.e.\ those
not well fit by a point source, single lens model without microlensing parallax)
from the MOA alert pages (https://it019909.massey.ac.nz/moa/) from the 
2007-2010 seasons. A similar analysis of OGLE-III binary events was conducted
by \citet{jar10}. The light curve calculations for this systematic analysis were 
done using the image centered ray-shooting method \citep{bennett96,bennett-himag}.
All the events that have anomalous deviation from a point source, single lens model 
were fitted with the binary lens model according to the following procedure.
Binary lens models have three parameters that are in common with single lens
models. These parameters are the time of the closest approach to the lens
system center of mass $t_0$, the Einstein radius crossing time, $t_{\rm E}$, 
and the impact parameter $u_0$, in units of the angular Einstein radius $\theta_{\rm E}$.
Binary lens models also require three additional parameters.
These are the lens system planet-star mass ratio, $q$, the planet-star separation, 
$s$, projected into the plane perpendicular to the line-of-sight, and 
the angle of the source trajectory relative to the binary lens axis $\alpha$.
Another parameter, the source radius crossing time, $t_\ast$ 
is also included in the binary lens model because this parameter is important for 
most of planetary microlensing light curves. 
We carefully searched to find the best fit binary lens model over a wide range of 
values of microlensing parameters using a variation of the Markov Chain Monte 
Carlo (MCMC) algorithm \citep{ver03}.
Because the shape of anomaly features in the light curve well depends on $q$, $s$ and $\alpha$, 
these three parameters are fixed in the first search.
Next, we searched $\chi^{2}$ minima of the 100 models in order of $\chi^{2}$ in the first search 
with each parameter free and found the best fit model.

\begin{figure}
\epsscale{0.70}
\centering
\includegraphics[width=120mm, angle=-90]{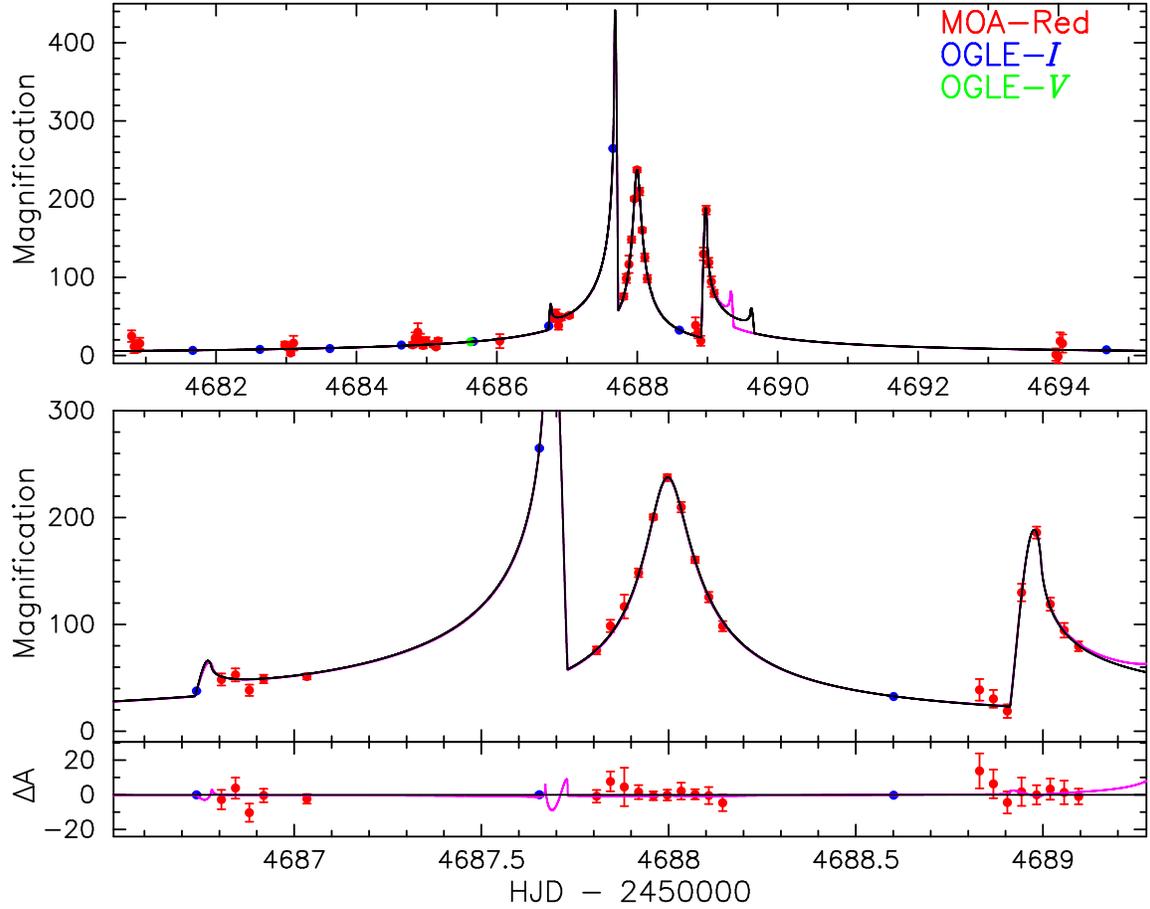}
\caption{Light curve of MOA-2008-BLG-379.
The top panel shows the magnified part of 
the light curve and the middle panel shows a close up of the anomaly.
The red, blue, and green points are for MOA-Red, OGLE-$I$, and OGLE-$V$, respectively. 
The black and magenta lines indicate the best close and wide model. 
The bottom panel shows the residual from the best close model.}
\label{light curve}
\end{figure}

This analysis indicated a planetary mass ratio for the MOA-2008-BLG-379L lens system.
Figure \ref{light curve} shows the light curve of this event and the best fit models.
There are five distinct caustic crossing and cusp approach features in the light curve. 
The first caustic crossing is sandwiched between a single observation by OGLE and
five MOA observations beginning 96 minutes later at ${\rm HJD}^\prime = 4686.8$,
The subsequent, very bright, caustic exit was observed with a single OGLE observation
on the subsequent night, but the central cusp approach feature and the next caustic
entrance feature were reasonably well sampled by 9 and 7 MOA observations, respectively,
on the next two nights. It is the sampling of these two features that allow the
parameters of the planetary lens model for this event to be determined.

For many binary events and most planetary events, the light curve has very sharp
features due to caustic crossings or cusp approaches that resolve the finite size of
the source star. Such features require an additional parameter, the source radius crossing
time, $t_\ast$. Since the MOA-2008-BLG-379
includes caustic crossings and a close cusp approach, it is necessary to include
finite source effects in its light curve model, and a proper accounting of finite
source effects requires that limb darkening be included.

For the limb darkening coefficients, we adopted a linear limb-darkening model 
for the source star based on the source color estimate of $(V-I)_{\rm S, 0}=0.81\pm0.13$,
which is discussed below in Section~\ref{sec-color}. This color
implies that the source is a late G-star, with 
an effective temperature of $T_{\rm eff} \simeq 5386\rm \ K$ \citep{bb88}.
We use limb-darkening parameters from \citet{clrt00} for 
a source star with an effective temperature $T_{\rm eff}=5500 \rm \ K$,  
surface gravity $\log g =4.5\rm\ cm\ s^{-2}$, and metallicity $\log {[\rm M/H]}=0.0$.
Girardi's isochrones \citep{girardi02} suggest that the source may be a metal poor star if it is
located at the distance of the Galactic center, but 
$\log {[\rm M/H]}=0.0$ is consistent within the 1-$\sigma$ error bar.
The list of the coefficients used for the linear limb-darkening model are as  
follows, 0.7107 for $V$, 0.5493 for $I$, and 0.5919 for MOA-Red, which is the 
mean of the coefficients for the Cousins $R$ and $I$-bands.
These are listed in Table \ref{info}.

\begin{figure}
\subfigure{
\includegraphics[width=75mm, angle=-90]{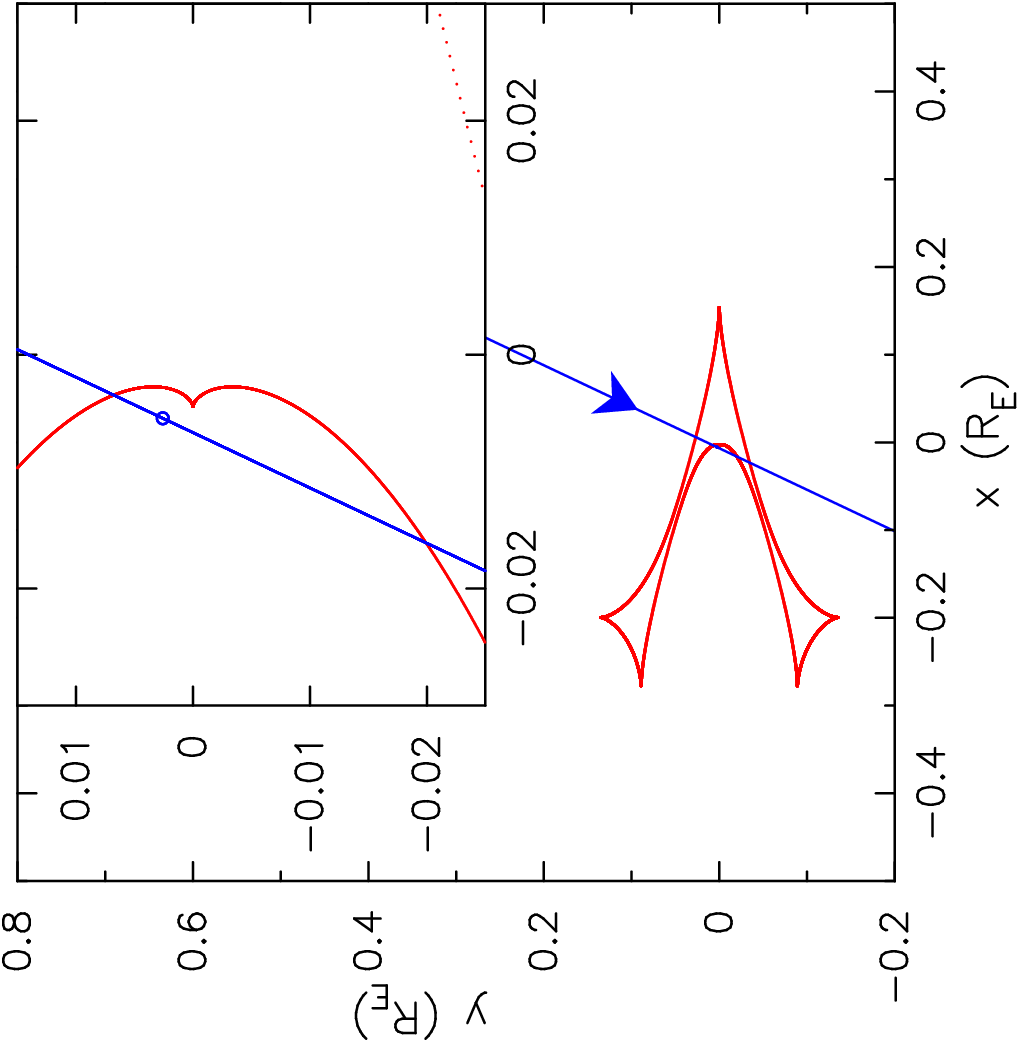}}
\subfigure{
\includegraphics[width=75mm, angle=-90]{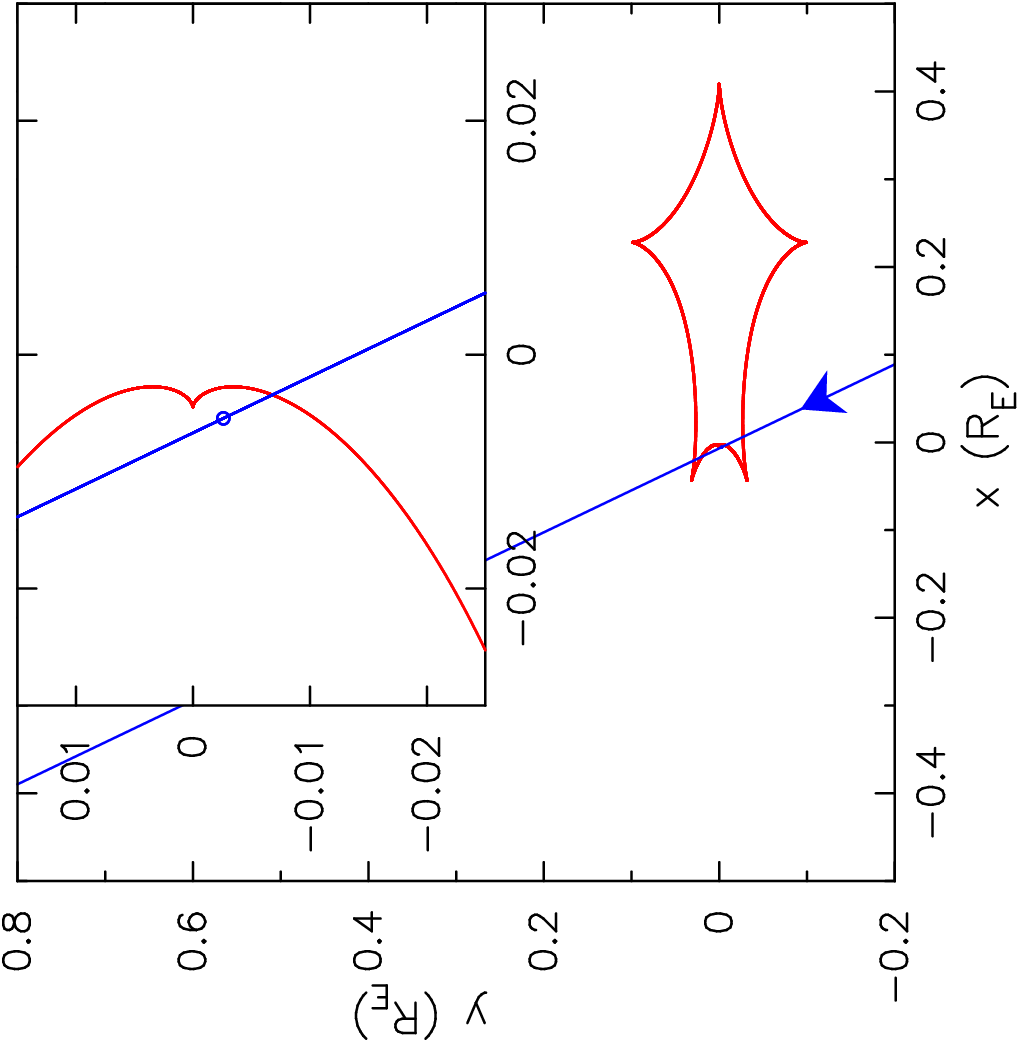}}

\caption{Caustic geometries for both close (left) and wide (right) models
are indicated by the red curves.
The insets in each panel are close up around the center of the coordinate.
The blue lines show the source star trajectory with respect to the lens system,
with arrows indicating the direction of motion.
The small blue circles in the insets indicate the source star size.}
\label{caustic}
\end{figure}

As is commonly the case for high magnification events, there are two 
degenerate light curve solutions that can explain the observed light curve.
This is the well-known ``close-wide" degeneracy, where the solutions have 
nearly identical parameters except that the separation is modified by
the $s \leftrightarrow 1/s$ transformation.
Figure \ref{caustic} shows the two caustic configurations from the close and wide 
models for this event. Sometimes, when the planetary caustics have merged with the
central caustic to form a so-called resonant caustic curve, as in this case, the light
curve data can resolve this close-wide degeneracy. We can see in Figure~\ref{light curve}
that the close and wide light curves do have substantial differences, such as the
time of the final caustic exit. However, the observations for this event are too sparse to 
sample these light curve features, and the degeneracy remains. Fortunately,
the parameters for these two solutions differ by only $\sim 20$\%, 
so this degeneracy has
little effect on the inferred physical parameters of the lens system. We find
$s=0.903\pm0.001$, $q=(6.85\pm0.05)\times10^{-3}$ and $u_0=(6.02\pm0.06)\times10^{-3}$
for the close model, which is preferred by $\Delta \chi^{2} = 0.7$, and
$s=1.119\pm0.001$, $q=(6.99\pm0.01)\times10^{-3}$, 
and $u_0=(6.03\pm0.03)\times10^{-3}$ for the wide model. The full set
of fit parameters are listed in Table~\ref{params}.

Higher order effects such as microlens parallax, xallarap (source star orbital motion)
and orbital motion in the lens system have been detected in some previous 
events\citep{alck95,gau08,bennett-ogle109,sumi10,mrk11,kains13}.
The measurement of finite source effects or microlensing parallax effects can 
partially break the degeneracies of the physical parameters that can be
inferred from the microlensing light curve, and the measurement of both
microlensing parallax and finite source effects yields a direct measurement
of the lens system mass \citep{bennett_rev,gaudi_araa}.
For this event, however, the source is too faint for a reliable microlensing
parallax measurement for an event of its duration, and the relatively sparse
data sampling leaves some uncertainty in the measurement of the
source radius crossing time. Therefore, we use the light curve constraints
on the source radius crossing time to constrain the physical parameters of the
lens system using a Bayesian analysis based on a standard Galactic model,
as discussed in Section \ref{sec-bayes}. If the lens star can be detected in
high angular resolution follow-up observations, it will then be possible to
directly determine the physical parameters of the lens system
\citep{bennett06,bennett07,dong-ogle71}.

\section{Source Color and Magnitude}
\label{sec-color}

\begin{figure}
\epsscale{0.80}
\centering
\includegraphics[width=120mm, angle=-90]{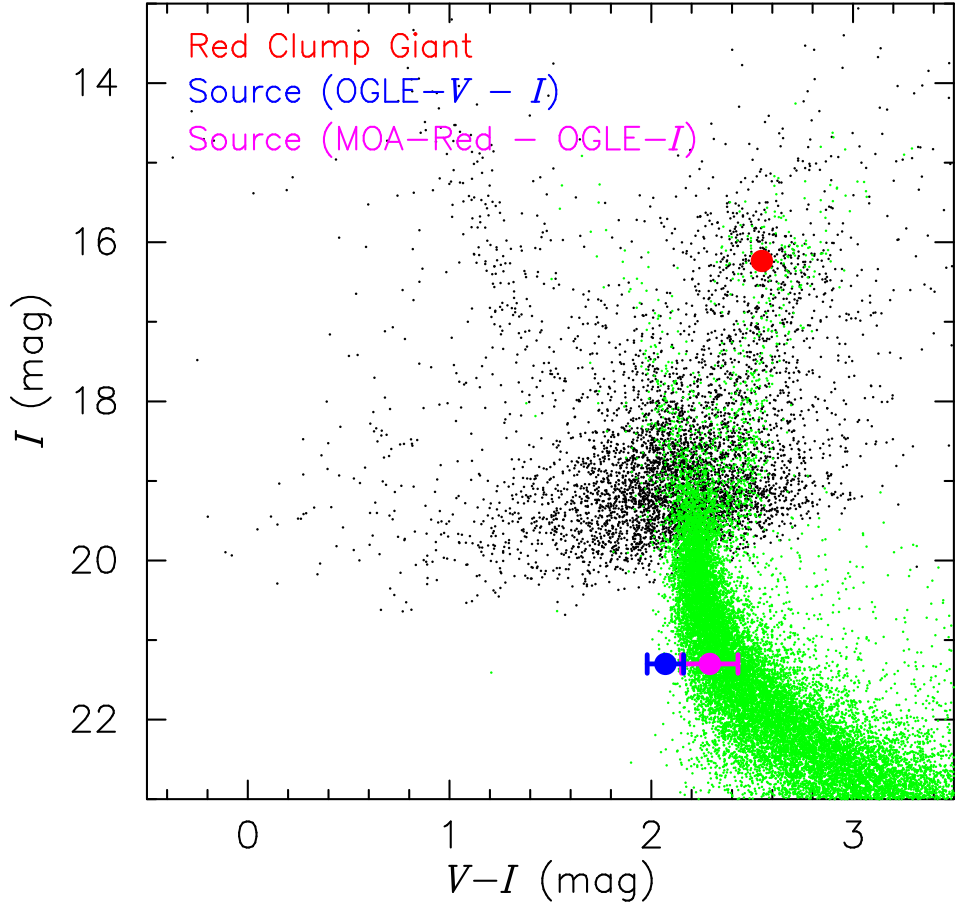}
\caption{Color--magnitude diagram (CMD) of the stars within $2^{\prime}$ of 
MOA-2008-BLG-379 from the OGLE-III catalog is shown as black dots.
The green dots show the $\it{HST}$ CMD of \citet{holtz98} whose 
extinction is adjusted to match the MOA-2008-BLG-379 using the Holtzman
field red clump giant (RCG) centroid of 
$(V-I, I)_{\rm RCG, Holtz} = (15.15, 1.62)$ \citep{ben08}. The red dot indicates
the RCG centroid for the MOA-2008-BLG-379 field.
The source star color and magnitude derived from OGLE $VI$ is indicated 
with a blue dot, while the source star color and magnitude derived from the
MOA-Red and OGLE-$I$ passbands is shown as a magenta dot. Although the
error bars on the MOA-Red + OGLE-$I$ color estimate are similar to the error
bar from the OGLE $VI$ estimate, we use the  MOA-Red + OGLE-$I$ 
estimate for the source color because it is subject to smaller systematic
uncertainties.}
\label{cmd_ogle}
\end{figure}

The source star magnitude and color estimated from the light curve modeling
are affected by extinction and reddening due to interstellar dust.
These effects must be removed in order to infer the intrinsic brightness and color of the source star.
In order to estimate extinction and reddening, we use the centroid of the red clump 
giant (RCG) distribution, which is an approximate standard candle.
The CMD, shown in Figure~\ref{cmd_ogle} 
was made from stars from the OGLE-III catalog \citep{szy11} within $2^\prime$
from the source star.
From this CMD, we have found the RCG centroid to be at
\begin{equation}
\label{eq:rcg}
(V-I,I)_{\rm RCG} = (2.55, 16.24)\pm(0.02, 0.04) \ .
\end{equation}

We adopt the intrinsic RCG centroid $V-I$ color \ and magnitude from 
\citet{bensby11} and \citet{nat13}, respectively, which gives
\begin{equation}
\label{eq:rcg0}
(V-I,I)_{\rm RCG,0} = (1.06, 14.42)\pm(0.06, 0.04) \ .
\end{equation}
Comparing the our measured RCG centroid from Equation~(\ref{eq:rcg})
with the intrinsic dereddened magnitude and extinction from 
Equation~(\ref{eq:rcg0}), we find that 
the reddening and extinction are 
\begin{equation}
\label{eq:ext_and_red}
(E(V-I),A_I)_{\rm RCG} = (1.49, 1.82)\pm(0.06, 0.06)
\end{equation}
, where $E(V-I)$ is the average reddening and $A_I$ is the average extinction.

The models presented in Table~\ref{params} give the source magnitude and
color of $I_{\rm S}=21.30\pm 0.03$ and $(V-I)_{\rm S}= 2.07\pm 0.09$ 
from the OGLE observations, calibrated to the OGLE-III photometry map \citep{szy11}. 
This color is bluer than most of the bulge main sequence stars at this
magnitude, but with the error bars, it is  marginally consistent with a bulge
main sequence star, as shown in Figure~\ref{cmd_ogle}.
This could be due to the fact that there is only a single bright OGLE $V$-band
image that might be affected by a nearby bright variable star. 

Because the MOA-Red passband is centered at a slightly shorter wavelength
than the OGLE $I$-band, we can use the MOA-Red and OGLE $I$ 
passbands to derive the source color \citep{gould10a}. Although the
color difference between these two passbands is relatively small, we
have a large number of data points at significant magnification in 
both of these passbands. So, this method should yield a  
color that is less sensitive to systematic photometry errors
than the determination based on the single magnified OGLE $V$-band measurement.
In order to calibrate the MOA-Red measurements to the OGLE-III ($V$, $I$) system,
we use the OGLE-III photometry map \citep{szy11} and a DoPHOT \citep{dophot}
reduction of the MOA reference frame. Because the seeing in the MOA reference
frame is significantly worse than the seeing images used for the OGLE-III catalog,
we choose isolated OGLE stars for the comparison with MOA to avoid problems
due to blending in the MOA images.
In order to select the stars with small uncertainty in magnitude and color, we 
remove faint stars with $I > 19$ and stars with a $V-I$ error bar $> 0.1$. 
In order to obtain an accurate linear relation, we only fit stars in the color
range $2 < V-I < 4$. We recursively reject $2.5 \sigma$ outliers and find
\begin{equation}
R_{M, \rm DoPHOT}  - I_{\rm OGLE-III} = (0.18931 \pm 0.00533) (V-I)_{\rm OGLE-III}
\label{eq-Rm-Io}
\end{equation}
, where $I_{\rm OGLE-III}$ and $V_{\rm OGLE-III}$ are the OGLE-III catalog magnitudes,
and $R_{M, \rm DoPHOT}$ is the calibrated MOA-Red band magnitude. (The MOA-Red calibration
has a zero-point uncertainty of 0.0144 mag.) 
The MOA light curve photometry \citep{bond01} is done with the difference imaging analysis
(DIA; \citealt{tom96,ala98}), which is usually significantly more precise than DoPHOT, but DIA only measures
changes in brightness, which is why DoPHOT photometry is used to derive the relation in
Equation~\ref{eq-Rm-Io} between the $R_{M}  - I_{\rm OGLE}$ and standard $V-I$ colors.
Thus, it is important that the MOA-Red band photometry from the DIA and DoPHOT codes
have identical magnitude scales. This is accomplished by using the DoPHOT point-spread 
function for the DIA photometry. Using the source magnitudes from the light curve modeling,
this procedure yields $(V-I, I)_{\rm S} = (2.29, 21.30)\pm(0.14, 0.03)$ for the best fit model.
As shown in Figure \ref{cmd_ogle}, this source magnitude and color are more reasonable for
a bulge source than the less precise color derived from OGLE $V$-band measurements, 
and it is the one that we use in our analysis.

Combining the extinction from Equation~(\ref{eq:ext_and_red}) with the source 
magnitudes and colors from the best fit wide and close models, we have
\begin{eqnarray}
\label{eq:VIIsw}
(V-I,I)_{\rm S,0} &=& (0.79, 19.48)\pm(0.13, 0.06)\ \rm for\ the\ wide\ model \\
(V-I,I)_{\rm S,0} &=& (0.80, 19.49)\pm(0.12, 0.06)\ \rm for\ the\ close\ model.
\label{eq:VIIsc}
\end{eqnarray}

Also, from the light curve models, we have determined the magnitude of the unlensed flux at the location of the source. 
This is the flux of the stars with images blended with the source. 
The flux from the blend stars is often used as an upper limit for the magnitude of the lens star.
In this event, however, we found that the blending flux in the light curve is dominated by the flux from a bright star $0.^{\prime \prime}5$ away from the event by the careful inspection of the OGLE reference image.
We found no star is resolved at the position of the event and thus the real upper limit of the lens magnitude is fainter.
If we assume that the extinction to these blend stars are the same as the source (which is a reasonable approximation if the stars are at a distance of $\simgt 4\,$kpc), 
 then we have the extinction-free 3-$\sigma$ limit of blend magnitude
\begin{equation}
I_{\rm b, 0}> 17.87~\rm mag
\label{eq:Ib}
\end{equation}
, which will be used as an upper limit for the magnitude of the planetary host star in Section \ref{sec-bayes}.

\section{Angular Einstein Radius $\theta_{\rm E}$}
\label{sec-thetaE}

With the use of $(V-I, I)_{\rm S,0}$, we can determine the source angular radius
with the color surface-brightness relation of \citet{kerfou08},
\begin{equation}
\log_{10}(2\theta_{\ast}) = 0.4992+0.6895(V-I)-0.0657(V-I)^{2}-0.2V\ .
\end{equation}
This yields an angular source radius of $\theta_{\ast}=0.45\pm0.07\ \mu$as.
$\theta_{\ast}$ is also able to be derived from the method of \citet{ker04} with $(V-K)_{\rm S,0}$ estimated from the dwarf star color-color relations from \citet{bb88} 
, but the result is consistent with the above value.
We can combine this value of $\theta_{\ast}$ with the fit source radius 
crossing time, $t_\ast$, values
from the light curve models to determine the lens-source relative proper motion,
\begin{eqnarray}
\label{eq:mu_rel_w}
\mu_{\rm rel}= \frac{\theta_\ast}{t_\ast} = \frac{\theta_{\rm E}}{t_{\rm E}}&=
           &7.46\pm1.65\ \rm mas\ yr^{-1}~for~the~wide~model \\    
       &=&7.75\pm1.56\ \rm mas\ yr^{-1}\ for~the~close~model. \label{eq:mu_rel_c}   
\end{eqnarray}
Note that this is the relative proper motion
in the instantaneously geocentric inertial reference frame that moved with the
Earth when the event reached peak magnification.
The measurement of $t_\ast$ also yields the angular Einstein radius
\begin{eqnarray}
\theta_{\rm E}=\frac{\theta_{\ast}t_{\rm E}}{t_\ast}&=&0.86 \pm 0.19\ \rm mas\ for\ the\ wide\ model\\    
                                                                              &=&0.90 \pm 0.18\ \rm mas~for~the~close~model \ , 
\end{eqnarray}
which can be used to help constrain the lens mass. 

As we discuss in Section \ref{sec-conclude}, follow-up observations might 
be able to detect the lens separating the source and measure the lens-source
relative proper motion. However, this would be in the heliocentric reference frame
rather than the instantaneously geocentric inertial frame used for
Equations~(\ref{eq:mu_rel_w}) and (\ref{eq:mu_rel_c}). Fortunately, the follow-up
observations and light curve model provide enough information to convert
between these two reference frames.

For this paper, however, we have not been able to distinguish the wide and
close models, so to obtain our final predictions, we combine the values from 
the wide and close models with weights given by $e^{-\Delta \chi^{2}/2}$.
This gives an angular Einstein radius and lens-source relative proper motion of
\begin{eqnarray}
\theta_{\rm E} &=& 0.88~\pm0.19\ \rm mas \\
\mu_{\rm rel} &=&7.63~\pm1.61\ \rm mas\ yr^{-1} \ ,
\end{eqnarray} 
with the lens-source relative proper motion in the same
instantaneous geocentric inertial reference frame used for 
Equations~(\ref{eq:mu_rel_w}) and (\ref{eq:mu_rel_c}).
 
\section{Bayesian Analysis}
\label{sec-bayes}

A microlensing light curve for a single lens event normally has the
lens distance, mass, and angular velocity with respect to the source 
constrained by only a single measured parameter, the Einstein radius
crossing time, $t_{\rm E}$. But, like most planetary microlensing events,
MOA-2008-BLG-379 has finite source effects that allow $t_{\ast}$,
and therefore $\theta_{\rm E}$, to be measured. If we could have also measured
the microlensing parallax effect, we could determine the total lens system
mass \citep{bennett_rev,gaudi_araa}.

Without a microlensing parallax measurement, we are left with the relation
\begin{equation}
\label{eq:mass-dist}
\theta^2_{\rm E} = \kappa M \left(\frac{1}{D_{\rm L}}-\frac{1}{D_{\rm S}}\right)
\end{equation}
where $\kappa = 8.14\ {\rm mas}/M_{\odot}$, $M$ is the mass of the lens system,
and $D_{\rm S}$ is the distance to the source star. This can be interpreted as
a lens mass--distance relation, since $D_{\rm S}$ is approximately known.

\begin{figure}
\epsscale{0.90}
\includegraphics[width=120mm, angle=-90]{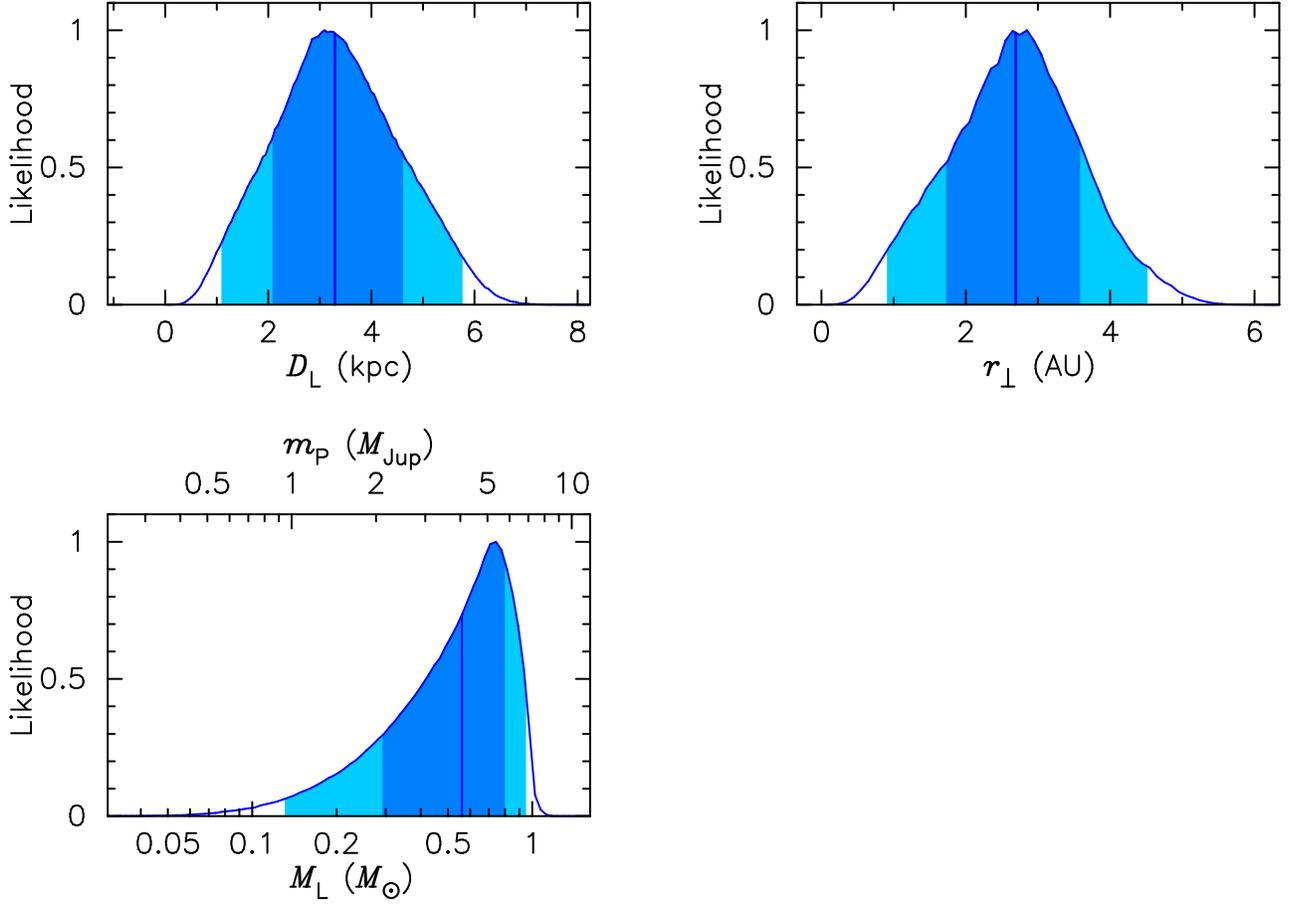}
\caption{Probability distribution of lens parameters from the Bayesian analysis.
The top-left panel shows the probability of the distance to the lens system.
The bottom-left panel shows the mass of the primary and 
secondary lenses (the star and planet) in units of Solar and Jupiter 
masses, respectively. 
The top-right panel shows the projected separation $r_{\perp}$.
The dark and light blue regions indicate the 68\% and 95\% confidence intervals, 
and the vertical lines indicate the median value.}
\label{like-out}
\end{figure}

We can now use this mass-distance relation, Equation~(\ref{eq:mass-dist}),
in a Bayesian analysis to estimate the physical properties of the lens system \citep{ogle390}.
Our Bayesian analysis used the Galactic model of \citet{hg03} with an assumed
Galactic center distance of 8 kpc with model parameters selected from the
MCMC used to find the best fit model. 
The lens magnitude is constrained to be less 
than the blend magnitudes presented in Equation~(\ref{eq:Ib}). 
Since the best fit wide model is slightly disfavored by $\Delta \chi^2  = 0.7$, we
weight the wide model Markov chains by $e^{-\Delta \chi^2 / 2}$ with
respect to the close model Markov chains. The probability distributions
resulting from this Bayesian analysis are shown in Figures \ref{like-out}
and \ref{like_filter}.

An important caveat to this Bayesian analysis is that we have assumed that
stars of all masses, as well as brown dwarfs, are equally likely to host a
planet with the measured mass ratio and separation. Because of this assumption,
the results of this Bayesian analysis cannot (directly) be used to determine
the probability that stars will host planets as a function of their mass.
With this caveat, the star and planet masses resulting from the Bayesian
analysis are $M_{\rm L} = 0.56_{-0.27}^{+0.24}\ M_{\odot}$ and
$m_{\rm P} = 4.1_{-1.9}^{+1.7}\ M_{\rm Jup}$, respectively.
Their projected separation was determined to be
$r_\perp = 2.7_{-1.0}^{+0.9}\ \rm AU$, and the lens system is at a 
distance of $D_{\rm L} = 3.3_{-1.2}^{+1.3}\ \rm kpc$.
The three-dimensional star-planet separation is estimated to be
$a = 3.3_{-1.2}^{+1.7}\ \rm AU$,
assuming a random inclination and phase. 
These values are listed in Table \ref{like_out}.
Therefore, the most likely physical parameters from the Bayesian analysis,
indicate that the planet has a mass of nearly 4 Jupiter-masses and orbits
a late K-dwarf host star at just over twice the distance of the snow line.

\section{Discussion and Conclusion}
\label{sec-conclude}

We reported on the discovery and analysis of a planetary microlensing event MOA-2008-BLG-379.
As is often the case with high magnification microlensing events, there are two
degenerate models: a close model with a planet-host separation of $s = 0.903$
and a wide model with $s = 1.119$. Both have a mass ratio of $q \simeq 7\times 10^{-3}$.
Our Bayesian analysis indicates that the lens system consists of a G, K, or 
M-dwarf orbited by a super-Jupiter mass planet. The most likely physical properties
for the lens system, according to the Bayesian analysis, are that the host is a late K-dwarf,
and the planet has a mass of about 4 Jupiter masses with a projected separation
of about $3\,$AU. However, these values are dependent on our prior assumption
that stars of different masses have equal probabilities to host a planet of the
observed mass ratio and separation. 

\begin{figure}
\epsscale{0.90}
\includegraphics[width=120mm, angle=-90]{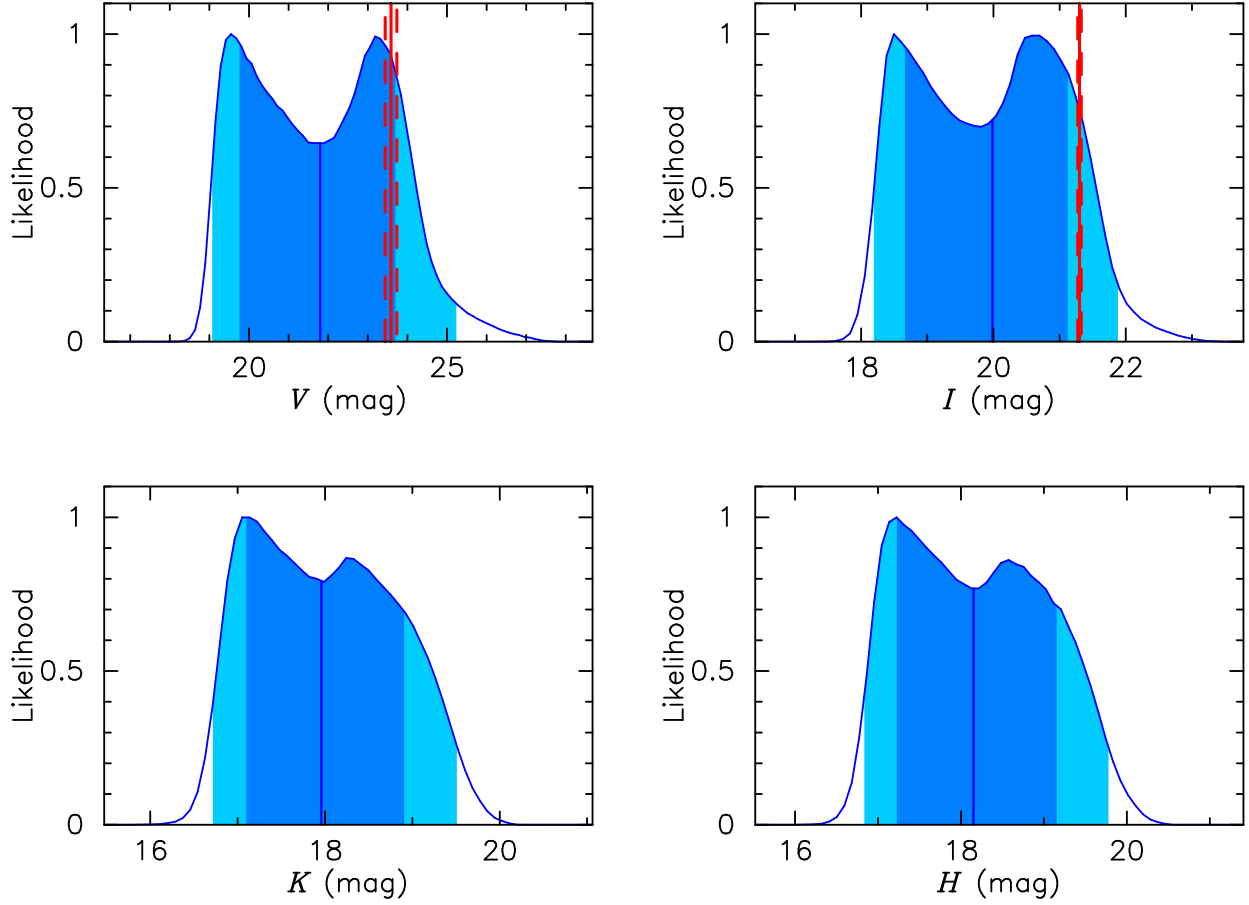}
\caption{Probability distribution of $V$-, $I$-, $K$- and $H$-band magnitudes for the extinction-free lens 
star from the Bayesian analysis.
The dark and light blue regions indicate the 68\% and 95\% confidence intervals.
The red solid lines show the source star magnitudes with the extinction estimated in Equation~(\ref{eq:ext_and_red}), and the red dashed lines are their 1 sigma errors. 
}
\label{like_filter}
\end{figure}

Fortunately, the parameters of this event are
quite favorable for a direct determination of the lens system mass and distance
by the detection of the lens separating from the source star in high angular
resolution follow-up observations \citep{bennett06,bennett07}. The source
star is quite faint at $I_{\rm S} = 21.30\pm 0.03$, so it is unlikely that the lens
will be very much fainter than the source. The brightness of the source is
compared to the Bayesian analysis predictions for the lens (host) star in
the $V$-, $I$-, $H$-, and $K$-bands in Figure \ref{like_filter}. This indicates
that the lens star is likely to have a similar brightness to the source in all
of these passbands. Also, it is unlikely that the lens and source magnitudes
will differ by more than 2 mag in any of these passbands, so both the source
and lens should be detectable.

The measured source radius crossing
time indicates a relatively large lens-source relative proper motion of 
$\mu_{\rm rel}=7.6\pm 1.6\ \rm mas\ yr^{-1}$ (measured in the inertial geocentric
reference frame moving at the velocity of the Earth at the peak of the event). This implies
a separation of $\sim 37.8\pm 8.3\,$mas in the later half of 2013 when high resolution
follow-up observations were obtained from Keck and the $\it Hubble~Space~Telescope$ ($\it HST$).
(Because the relative proper motion, $\mu_{\rm rel}$, is measured in an instantaneously 
geocentric reference frame, it cannot be directly converted into a precise separation
prediction, but the conversion to the heliocentric reference frame, which is more relevant to the
follow-up observation, usually results in a small change.)
The results of these follow-up observations will be reported in a future publication, but
it seems quite likely that the planetary host star will be detected, resulting in
a complete solution of the lens system. That is, the lens masses, distance and
separation will all be determined in physical units.

One unfortunate feature of this discovery is that MOA-2008-BLG-379 
was not recognized as a planetary microlensing event when it was observed. This
was partly due to the faintness of the source star, which rendered the relatively long
planetary perturbation as the dominant portion of the magnified light curve.
However, the lack of familiarity with the complete variety of binary lens and planetary
microlensing light curves also played a role in this lack of recognition.
The planetary nature of the event was discovered as a result of a systematic effort
to model all of the binary microlensing events that were found by the MOA alert
system.
Fortunately, the joint analysis of OGLE data and the high cadence MOA-II survey observations
allowed the planetary nature of the event to be establish and the basic planetary lens parameters 
to be measured.

The MOA-II high cadence survey was enabled by the wide, $2.2\, {\rm deg}^2$
field MOA-cam3 that enabled hourly survey observations. 
In 2010 March, the OGLE group started the OGLE-IV survey with 
their new $1.4\, \rm deg^2$ CCD camera, and they now also conduct a high cadence
survey of the central Galactic bulge. This new camera and the good seeing available
at the OGLE site (Las Campanas) has substantially increased the rate
in which planetary deviations are discovered in survey observations. The 
combination of the MOA-II and OGLE-IV high cadence surveys greatly improves
the light curve sampling of these survey microlensing planet discoveries due to
the large separation in longitude between the MOA and OGLE telescopes.
The capabilities of microlensing follow-up teams are also rapidly 
improving, most notably with the recent expansion \citep{lcogt} of the 
Las Cumbres Observatory Global Telescope Network (LCOGT), which 
should result in much higher light curve sampling rates of planetary signals
discovered in progress.



\acknowledgments
We acknowledge the following sources of support: the MOA project was 
supported by the Grant-in-Aid for Scientific Research (JSPS19015005, 
JSPS19340058, JSPS20340052, JSPS20740104).
D.S.\ was supported by Grant-in-Aid for JSPS Fellows.
D.P.B.\ was supported by grants
NASA-NNX12AF54G, JPL-RSA 1453175 and NSF AST-1211875.
The OGLE project has received funding from the European Research Council under the European Community's Seventh Framework Programme (FP7/2007-2013)/ERC grant agreement No. 246678 to A.U.





\appendix




\clearpage



\clearpage


\clearpage

\begin{table}
\begin{center}
\caption{The error bar corrections parameters and linear limb darkening parameters
for the data sets used to model the MOA-2008-BLG-379 light curve.
The formula used to modify the error bars is 
$\sigma^{'}_i=k\sqrt{\sigma^2_i+e^2_{\rm min}}$ where $\sigma_i$ is the 
input error bar for the $i$th data point from the photometry code in 
magnitudes, and $\sigma^{'}_i$ is the final error bar used for the
determination of parameter uncertainties.\newline}
\label{info}
\begin{tabular}{lcccc}
\tableline \tableline
Dataset & $k$ & $e_{\rm min}$ & Limb-darkening Coefficients & Number of Data \\ \tableline
MOA-Red & 1.266 & 0 & 0.5919 & 951  \\
OGLE $I$-band & 0.995 & 0 & 0.5493 & 294  \\
OGLE $V$-band & 1.180 & 0 & 0.7107 & 7  \\ \tableline
\end{tabular}
\end{center}
\end{table}

\clearpage

\begin{table}
\begin{center}
\caption{The best fit model parameters for both the wide and close 
models. The second and fourth rows in each column are the 1-$\sigma$ error bars
for each parameter. $\rm HJD^\prime \equiv HJD-2450000$. \newline}
\label{params}
\begin{tabular}{l|ccccccccc}
\tableline \tableline
Model & $t_0$ & $t_{\rm E}$ & $u_0$ & $q$ & $s$ & $\theta$ & $t_\ast$ & $\chi^2$ \\
 & $\rm HJD^\prime$ & (days) & $10^{-3}$ & $10^{-3}$ & & (rad) & (days) & \\ \tableline
Wide  & 4687.896 & 42.14 & 6.03 & 6.99 & 1.119 & 1.124 & 0.0219 & 1246.7 \\ 
         & 0.001        & 0.23 & 0.03 & 0.01 & 0.001 & 0.003 & 0.0020 \\
Close & 4687.897 & 42.46 & 6.02 & 6.85 & 0.903 & 5.154 & 0.0212 & 1246.0 \\   
         & 0.001        & 0.45   & 0.06 & 0.05 & 0.001 & 0.002 & 0.0017 & \\ \tableline 
\end{tabular}
\end{center}
\end{table}





\clearpage

\begin{table}
\caption{Physical parameters of the lens system obtained from the Bayesian analysis.}
\label{like_out}
\begin{center}
\begin{tabular}{ccccc}
\tableline \tableline
Primary Mass & Secondary Mass & Distance & Projected Separation & Separation \\
 $M_{\rm L}\ (M_{\odot})$ & $m_{\rm P}\ (M_{\rm Jup})$ & $D_{\rm L}\ (\rm kpc)$ & $r_{\perp}\ (\rm AU)$ & $a\ (\rm AU)$ \\ \tableline
$0.56_{-0.27}^{+0.24}$ & $4.1_{-1.9}^{+1.7}$ & $3.3_{-1.2}^{+1.3}$ & $2.7_{-1.0}^{+0.9}$ & $3.3_{-1.2}^{+1.7}$ \\ \tableline 
\end{tabular}
\end{center}
\end{table}





\end{document}